\documentstyle[preprint,aps,12pt]{revtex}
\begin{document}
\draft
\title{Spin-ordering and magnon collective modes for two-dimensional electron
lattices in strong magnetic fields}
\author{R. C\^ot\'e}
\address{D\'epartement de Physique et Centre de Recherches en\\
Physique du Solide, Universit\'e de Sherbrooke, Sherbrooke,\\
Qu\'ebec, Canada J1K-2R1}
\author{A. H. MacDonald}
\address{Department of Physics, Indiana University, Bloomington, Indiana 47405}
\date{\today}
\maketitle

\begin{abstract}
We study the spin-ordering and the magnon collective modes of the
two-dimensional Wigner crystal state at strong magnetic fields. Our work is
based on the Hartree-Fock approximation for the ground state and the
time-dependent Hartree-Fock approximation for the collective modes. We find
that the ground state is ferromagnetic, {\it i.e.,} that all spins are
aligned at $T=0$ even when the electronic g-factor is negligibly small. The
magnon calculations show that the spin-stiffness is much smaller in the
crystal state than in fluid states which occur at nearby Landau level
filling factors.
\end{abstract}

\pacs{PACS numbers: 73.20.Mf, 75.30.Ds, 75.30.Et}

\section{INTRODUCTION}

At sufficiently low electron densities or in sufficiently strong magnetic
fields, electrons will crystallize at low temperatures.\cite{wigner,chui} It
is generally expected that in the (Wigner) crystal state, electronic spins
will be either ferromagnetically or antiferromagnetically ordered.\cite
{fieldcaveat,gfac} There has long been theoretical interest\cite
{herring,thouless} in the rather subtle physics which determines how the
electronic spins are ordered. For two-dimensional electrons in the Wigner
crystal state at zero magnetic field a series\cite{ceperley,louie} of
variational and Green's function Monte Carlo calculations have not led to
definitive conclusions concerning the nature of the magnetic order.{\bf \ }%
The energetically preferred spin ordering has been shown to depend very much
on the lattice structure of the electron crystal and, unfortunately, the
difference in energy between states with different spin order on a hexagonal
lattice (which is expected to be the ground-state lattice of the Wigner
crystal) is smaller than the accuracy of the Monte Carlo calculations\cite
{louie}. To our knowledge, there are no previous numerical studies of the
spin structure of the Wigner crystal in the strong field regime. The
variational Monte Carlo calculations cited above considered, in the
strong-field limit, only the spin-polarized hexagonal lattice and
investigated exchange, correlation and Landau-level-mixing effects\cite
{louie}.

In this paper we discuss magnetic order for two-dimensional electrons in the
limit of strong perpendicular magnetic fields where all electrons are
confined to the lowest quantized kinetic energy Landau level. In this limit,
the state of the electrons depends on the Landau level filling factor $\nu $
rather than the electron density and, except for a narrow interval
surrounding $\nu =0.2$, the electrons form a Wigner crystal state\cite{chui}
for $\nu $ smaller than $\approx 0.23$. ($\nu \equiv N/N_\phi $ where $N$ is
the number of electrons and $N_\phi =SB/\Phi _0\equiv S/(2\pi \ell ^2)$ is
the Landau level degeneracy. Here $S$ is the area of the system, $B$ is the
magnetic field strength, $\ell $ is the magnetic length and $\Phi _0=hc/e$
is the magnetic flux quantum.) We find, partly on the basis of Hartree-Fock
approximation (HFA) calculations, that in this regime the Wigner crystal
state will always be ferromagnetic. The Hartree-Fock ground-state wave
function does not contain the important correlations that give rise to the
magneto-phonon and spin waves modes of the crystal. We can check, however,
the stability of the spin-polarized lattice by evaluating the magnon
spectrum of the Wigner crystal using a time-dependent Hartree-Fock
approximation (TDHFA). Within the limits of our numerical approach, we find
that the polarized lattices remain stable at small filling factors.
Moreover, in this limit, the spin-wave modes are very well described by a
Heisenberg model where electrons are localized on their lattice site with an
effective exchange integral $J_{TDHFA}$ that we compute for different
filling factors. From the value of this effective exchange integral, we can
derive the spin-stiffness of the Wigner crystal state and compare its value
with the spin-stiffness of ferromagnetic electron fluid states at nearby
filling factors. This comparison shows that the spin-stiffness of the liquid
is much larger than that of the solid. This is so because, the exchange
energy is larger when the electrons are free to move around and come closer
to each other.

Our paper is organized as follows. In Section II we outline the formalism we
use to perform the HFA and TDHFA calculations calculations for respectively
the ground state and spin waves of the system. We are able to enormously
simplify the calculations by adapting an approach we developed\cite{cote1}
previously to the situation of interest here. In Section III we present
numerical results for the magnetic ground state of the square and triangular
Wigner crystal states and discuss differences between magnetic ordering
tendencies in zero-field and strong-field limits. Our results for magnon
dispersion relation are presented and discussed in Section IV. Section V
contains a brief summary of this work.

\section{Hartree-Fock and Time-Dependent Hartree-Fock Approximations}

\subsection{Hartree-Fock Approximation Ground State}

The formalism outlined in this section is a straightforward generalization
of one which we developed originally\cite{cote1} to describe phonon modes in
the Wigner crystal state and which has previously been generalized in other
directions\cite{cote2,others} to describe double-layer quantum Hall systems
and edge excitations of the Wigner crystal. We outline the main steps in the
development of the formalism and refer the reader to Ref. \cite{cote1} for
further details.

We consider a two-dimensional electron gas (2DEG) in a magnetic field ${\bf B%
}=-B\widehat{{\bf z}}$ which is assumed to be strong enough so that we can
make the usual approximation of considering only the lowest Landau level. In
the Landau gauge, the Hamiltonian of the 2DEG is then (we set $\hbar =1$
throughout this paper)
\begin{eqnarray}
H=\sum_{\alpha ,X}\epsilon _\alpha c_{\alpha ,X}^{\dagger }c_{\alpha ,X} &+&%
\frac 1{2S}\sum_{{\bf q}}\sum_{{X_1,...,X_4}}\sum_{\alpha ,\beta }V({\bf q}%
)\langle X_1|\exp (i{\bf q}\cdot {\bf r})|X_4\rangle  \nonumber \\
&&\quad \quad \quad \times \langle X_2|\exp (-i{\bf q}\cdot {\bf r}%
)|X_3\rangle c_{\alpha ,X_1}^{\dagger }c_{\beta ,X_2}^{\dagger }c_{\beta
,X_3}c_{\alpha ,X_4},  \label{un}
\end{eqnarray}
where $\alpha ,\beta ={+}$(up), ${-}$ (down) are spin indices and the lowest
Landau level has the energy
\begin{equation}
\epsilon _\alpha =\frac{\omega _c}2-\frac{\alpha g^{*}\mu _bB}2.
\label{deux}
\end{equation}
As usual, $\omega _c=eB/m^{*}c$ is the cyclotron frequency and $m^{*}$ and $%
g^{*}$ are the effective mass and $g$-factor of the electron appropriate\cite
{gfac} to the two-dimensional electron layer. For a finite system, the
allowed values of the quantum number $X$ are separated by $2\pi \ell ^2/L_y$%
. Neglecting the finite thickness of the two-dimensional electron layer, we
take $V({\bf q})=2\pi e^2/q$, the two-dimensional Fourier transform of the
Coulomb potential.

Making the usual Hartree-Fock pairing of the second-quantized operators in
the Hamiltonian of Eq. (\ref{un}) and allowing for the possibility of broken
translational symmetry and spin magnetic order in the ground-state, we
obtain
\begin{equation}
H=N_\phi \sum_\alpha \epsilon _\alpha \rho _{\alpha ,\alpha }(0)+N_\phi
\sum_{{\bf q}}\sum_{\alpha ,\beta }V_{\alpha ,\beta }({\bf q})\rho _{\alpha
,\beta }({\bf q}),  \label{trois}
\end{equation}
where we have introduced the operators
\begin{equation}
\rho _{\alpha ,\beta }({\bf q})=N_\phi ^{-1}\sum_X\exp \left(
-iq_xX-iq_xq_y\ell ^2/2\right) c_{\alpha ,X}^{\dagger }c_{\beta ,X+q_y\ell
^2},  \label{quatre}
\end{equation}
which are related to the density and spin operators by the relations
\begin{equation}
\begin{array}{rcl}
n(q) & = & N_\phi e^{-q^2\ell ^2/4}\left[ \rho _{++}({\bf q})+\rho _{--}(%
{\bf q})\right] , \\
S^z(q) & = & \frac 12N_\phi e^{-q^2\ell ^2/4}\left[ \rho _{++}({\bf q})-\rho
_{--}({\bf q})\right] , \\
S^{+}(q) & = & N_\phi e^{-q^2\ell ^2/4}\rho _{+-}({\bf q}), \\
S^{-}(q) & = & N_\phi e^{-q^2\ell ^2/4}\rho _{-+}({\bf q}).
\end{array}
\label{cinq}
\end{equation}
The matrix elements of the Hartree-Fock self-consistent field in Eq.(\ref
{trois}) are given by
\begin{equation}
\begin{array}{rcl}
V_{++}({\bf q}) & = & \left[ H({\bf q})-X({\bf q})\right] \left\langle \rho
_{++}(-{\bf q})\right\rangle +H({\bf q})\left\langle \rho _{--}(-{\bf q}%
)\right\rangle , \\
V_{--}({\bf q}) & = & \left[ H({\bf q})-X({\bf q})\right] \left\langle \rho
_{--}(-{\bf q})\right\rangle +H({\bf q})\left\langle \rho _{++}(-{\bf q}%
)\right\rangle , \\
V_{+-}({\bf q}) & = & -X({\bf q})\left\langle \rho _{-+}(-{\bf q}%
)\right\rangle , \\
V_{-+}({\bf q}) & = & -X({\bf q})\left\langle \rho _{+-}(-{\bf q}%
)\right\rangle ,
\end{array}
\label{six}
\end{equation}
with the Hartree ($H$) and Fock ($X$) interactions defined by

\begin{equation}
\begin{array}{rcl}
H({\bf q}) & = & {\left( \frac{e^2}\ell \right) }\left( \frac 1{q\ell }%
\right) e^{-q^2\ell ^2/2}\,(1-\delta _{{\bf q},0}), \\
X({\bf q}) & = & \left( \frac{e^2}\ell \right) \sqrt{\frac \pi 2}e^{-q^2\ell
^2/4}I_0(q^2\ell ^2/4),
\end{array}
\label{sept}
\end{equation}
where $I_0(x)$ is the modified Bessel function of the first kind and the
factor $(1-\delta _{{\bf q},0})$ comes from the neutralizing positive
background.

The ordered state is defined by the set of order parameters $\left\{
\left\langle \rho _{\alpha ,\beta }({\bf q})\right\rangle \right\} .$ In the
case of interest here, {\em i.e.} for a Wigner lattice, these parameters are
non-zero only when ${\bf q}={\bf G},$ a reciprocal lattice vector of the
crystal. To calculate the $\left\langle \rho _{\alpha ,\beta }({\bf q}%
)\right\rangle $ 's, we define the $2\times 2$ single-particle Green's
function
\begin{equation}
G_{\alpha ,\beta }(X,X^{\prime },\tau )=-\left\langle Tc_{\alpha ,X}(\tau
)c_{\beta ,X^{\prime }}^{\dagger }(0)\right\rangle ,  \label{huit}
\end{equation}
and its Fourier transform $G_{\alpha ,\beta }({\bf q},\tau )$ by
\begin{equation}
G_{\alpha ,\beta }({\bf q},\tau )=N_{\phi} ^{-1}\sum_{X,X^{\prime
}}G_{\alpha ,\beta }(X,X^{\prime },\tau )\exp \left[ -\frac 12%
iq_x(X+X^{\prime })\right] \delta _{X^{\prime },X-q_y\ell ^2},  \label{neuf}
\end{equation}
so that
\begin{equation}
\langle \rho _{\alpha ,\beta }({\bf q})\rangle =G_{\beta ,\alpha }({\bf q}%
,\tau =0^{-}).  \label{dix}
\end{equation}

Using the Heisenberg equation of motion $\frac \partial {\partial \tau }%
(\ldots )=[H-\mu N,(\ldots )]$ where $\mu $ is the chemical potential of the
electrons which we measure with respect to the kinetic energy of the first
Landau level, we obtain the equation of motion for the single-particle
Green's function (in an obvious matrix notation)
\begin{equation}
\left[ \left( i\omega _n+\mu \right) I-\Lambda \left(
\begin{array}{cc}
1 & 0 \\
0 & -1
\end{array}
\right) \right] G({\bf q},\omega _n)-\sum_{{\bf q}^{\prime }}\exp \left[
\frac 12i{\bf q}\times {\bf q}^{\prime }\ell ^2\right] V({\bf q}^{\prime }-%
{\bf q})G({\bf q}^{\prime },\omega _n)=I\delta _{{\bf q},0},  \label{dixp}
\end{equation}
where $\omega _n$ is a fermionic Matsubara frequency, $\Lambda \equiv
g^{*}\mu _bB/2$ and $I$ is the $2\times 2$ unit matrix.

Eq. (\ref{dixp}) is very general. For example, it can be used to consider
complex spin-texture states such as the Skyrme crystal studied in Ref. \cite
{cote2} where the average value of all three components of the average spin
are space dependent. Although we concentrate, in this work, on simple
spin-structure states where Eq. (\ref{dixp}) can be reduced to only one
uncoupled equation, we explain here our numerical approach for the general
case.

We represent by ${\bf q}_1,{\bf q}_2,{\bf q}_3,\ldots {\bf q}_N$ the wave
vectors defining the ordered state (in principle, $N\rightarrow \infty $ but
in the numerical calculation, a suitable cutoff is chosen for $N$). We
choose ${\bf q}_1=0$ and define the vector $\widetilde{G}_{\alpha ,\beta
}\equiv (G_{\alpha ,\beta }({\bf q}_1),G_{\alpha ,\beta }({\bf q}%
_2),G_{\alpha ,\beta }({\bf q}_3),\ldots ,G_{\alpha ,\beta }({\bf q}_N)).$
Since in Eq. (\ref{dixp}), $G_{++}$ (or $G_{--}$ ) is coupled to $G_{-+}\,$
(or $G_{+-}$ ) only, we can simplify Eq. (\ref{dixp}) by defining the $2N-$%
component vectors $\widetilde{G}_1\equiv (\widetilde{G}_{++},\widetilde{G}%
_{-+})$ and $\widetilde{G}_2\equiv (\widetilde{G}_{+-},\widetilde{G}_{--}).$
We finally get a set of two coupled integral equations that we write in
matrix form as

\begin{equation}
(i\omega _n+\mu )\widetilde{I}\left(
\begin{array}{c}
\widetilde{G}_{++} \\
\widetilde{G}_{-+}
\end{array}
\right) -\widetilde{F}\left(
\begin{array}{c}
\widetilde{G}_{++} \\
\widetilde{G}_{-+}
\end{array}
\right) =\left(
\begin{array}{c}
\widetilde{1} \\
\widetilde{0}
\end{array}
\right) ,  \label{onze}
\end{equation}
and
\begin{equation}
(i\omega _n+\mu )\widetilde{I}\left(
\begin{array}{c}
\widetilde{G}_{+-} \\
\widetilde{G}_{--}
\end{array}
\right) -\widetilde{F}\left(
\begin{array}{c}
\widetilde{G}_{+-} \\
\widetilde{G}_{--}
\end{array}
\right) =\left(
\begin{array}{c}
\widetilde{0} \\
\widetilde{1}
\end{array}
\right) .  \label{onzep}
\end{equation}
In these equations, $\widetilde{I}$ is the $2N\times 2N$ unit matrix, $%
\widetilde{1}\equiv (1,0,0,\ldots ,0)$ and $\widetilde{0}\equiv
(0,0,0,\ldots ,0)$ are respectively the $N-$component unit and nul vector,
and $\widetilde{F}$ is the $2N\times 2N\ $matrix defined by $\ $
\begin{equation}
\widetilde{F}\equiv \left[
\begin{array}{cc}
\Lambda \delta _{{\bf q},{\bf q}^{\prime }}+A_{++}({\bf q},{\bf q}^{\prime })
& A_{+-}({\bf q},{\bf q}^{\prime }) \\
A_{-+}({\bf q},{\bf q}^{\prime }) & -\Lambda \delta _{{\bf q},{\bf q}%
^{\prime }}+A_{--}({\bf q},{\bf q}^{\prime })
\end{array}
\right] ,  \label{douze}
\end{equation}
where
\begin{equation}
A_{\alpha ,\beta }({\bf q},{\bf q}^{\prime })=\exp \left[ \frac 12i{\bf q}%
\times {\bf q}^{\prime }\ell ^2\right] V_{\alpha ,\beta }({\bf q}^{\prime }-%
{\bf q}).  \label{treize}
\end{equation}

Note that since $A_{\alpha ,\beta }({\bf q},{\bf q}^{\prime })=\left[
A_{\beta ,\alpha }({\bf q}^{\prime },{\bf q})\right] ^{*},$ $\widetilde{F}$
is an hermitian matrix. It follows that Eqs. (\ref{onze}) and (\ref{onzep})
can be solved by making the unitary transformation $\widetilde{F}%
=UDU^{\dagger }$, where $UU^{\dagger }=1$ and $D$ is the diagonal matrix
containing the eigenvalues of $\widetilde{F}.$ Following Ref. \cite{cote1},
we have for the order parameters ( $i=1,2,\ldots ,N$ )
\begin{equation}
\begin{array}{rcl}
\left\langle \rho _{++}({\bf q}_i)\right\rangle & = & \sum_{k=1}^{k=k_{\max
}}\ U_{i,k}U_{1,k}^{*}, \\
\left\langle \rho _{+-}({\bf q}_i)\right\rangle & = & \sum_{k=1}^{k=k_{\max
}}\ U_{i+N,k}U_{1,k}^{*}, \\
\left\langle \rho _{-+}({\bf q}_i)\right\rangle & = & \sum_{k=1}^{k=k_{\max
}}\ U_{i,k}U_{N+1,k}^{*}, \\
\left\langle \rho _{--}({\bf q}_i)\right\rangle & = & \sum_{k=1}^{k=k_{\max
}}\ U_{i+N,k}U_{N+1,k}^{*}.
\end{array}
\label{qorze}
\end{equation}
The value of $k_{\max }$ is obtained from the conditions
\begin{equation}
\left\langle \rho _{++}(0)\right\rangle =\nu _{+}\;,  \label{quinze}
\end{equation}
and
\begin{equation}
\left\langle \rho _{--}(0)\right\rangle =\nu _{-}\;,  \label{seize}
\end{equation}
the filling factors for spin up and down. It is easy to show from Eq. (\ref
{qorze}) that, at $T=0K,$ the following sum rules hold
\begin{equation}
\sum_{{\bf q}}\left[ \left| \left\langle \rho _{++}({\bf q})\right\rangle
\right| ^2+\left| \left\langle \rho _{+-}({\bf q})\right\rangle \right|
^2\right] =\nu _{+},  \label{dsept}
\end{equation}
and
\begin{equation}
\sum_{{\bf q}}\left[ \left| \left\langle \rho _{-+}({\bf q})\right\rangle
\right| ^2+\left| \left\langle \rho _{--}({\bf q})\right\rangle \right|
^2\right] =\nu _{-}.  \label{dhuit}
\end{equation}

We note that, except for simple cases such as the fully polarized or
unpolarized crystals, the filling factors $\nu _{+}$ and $\nu _{-}$ are not
known from the beginning. The only boundary conditions are the constraints
\begin{equation}
\langle \rho _{++}(0)\rangle +\langle \rho _{--}(0)\rangle =\nu ,
\label{dneuf}
\end{equation}
and
\begin{equation}
\langle \rho _{+-}(0)\rangle =\langle \rho _{-+}(0)\rangle =0,\qquad {\rm %
if\qquad }\Lambda \neq 0.  \label{vingt}
\end{equation}
Also, by definition,

\begin{equation}
\langle \rho _{+-}({\bf q})\rangle =\langle \rho _{-+}(-{\bf q})\rangle ^{*}.
\label{vingtp}
\end{equation}
To find $\nu _{+}$ , $\nu _{-},$ Eqs. (\ref{onze},\ref{onzep}) must be
solved self-consistently for a given value of $\nu _{+}$ and $\nu _{-}$
until a convergent solution is obtained. The process has to be repeated for
different sets of $\nu _{+}$ and $\nu _{-}$ values until the lowest-energy
solution is found. In this way we can determine the lowest-energy single
Slater determinant consistent with any assumed translational and magnetic
symmetry. The Hartree-Fock energy per particle of a particular ground-state
configuration (with respect to the kinetic energy of the lowest Landau
level) is:

\begin{eqnarray}
E &=&\Lambda \left( \frac{\nu _{{-}}-\nu _{+}}\nu \right) +\frac 1{2\nu }%
\sum_{{\bf q}}\left\{ \left[ H({\bf q})-X({\bf q})\right] \left[ |\langle
\rho _{++}({\bf q})\rangle |^2+|\langle \rho _{--}({\bf q})\rangle
|^2\right] \right.  \label{vun} \\
\quad \quad \qquad &&+H({\bf q})\left[ \langle \rho _{++}({\bf q})\rangle
\langle \rho _{--}(-{\bf q})\rangle +h.c.\right] -\left. 2X({\bf q})|\langle
\rho _{+-}({\bf q})\rangle |^2\right\}  \nonumber
\end{eqnarray}

In the case of a fully spin-polarized Wigner crystal, only $\left\{ \langle
\rho _{{++}}({\bf G})\rangle \right\} \neq 0$ (${\bf G}$ is a reciprocal
lattice vector) and Eq. (\ref{dixp}) simplifies to
\begin{equation}
\left( i\omega _n+\mu -\Lambda \right) G_{{++}}({\bf G},\omega _n)-\sum_{%
{\bf G}^{\prime }}\widetilde{F}({\bf G},{\bf G}^{\prime })G_{{++}}({\bf G}%
^{\prime },\omega _n)=\delta _{{\bf G},0},  \label{vsept}
\end{equation}
where $\widetilde{F}\ $is now the $N\times N$ matrix
\begin{equation}
\widetilde{F}({\bf G},{\bf G}^{\prime })=\exp \left[ \frac 12i{\bf G}\times
{\bf G}^{\prime }\ell ^2\right] V_{{++}}({\bf G}-{\bf G}^{\prime }),
\label{vseptp}
\end{equation}
with
\begin{equation}
V_{{++}}({\bf G})=\left[ H({\bf G})-X({\bf G})\right] \left\langle \rho _{{++%
}}({\bf G})\right\rangle .  \label{vhuit}
\end{equation}
The ground-state energy per particle, in this case, is simply
\begin{equation}
E_{{++}}=-\Lambda +\frac 1{2\nu }\sum_{{\bf G}}\left[ H({\bf G})-X({\bf G}%
)\right] \left| \left\langle \rho _{{++}}({\bf G})\right\rangle \right| ^2.
\label{vneuf}
\end{equation}

\subsection{Time-Dependent Hartree-Fock Approximation Collective Excitations}

To determine the collective excitation energies of the ordered state, we
define the response functions
\begin{equation}
\chi _{\alpha \beta \gamma \delta }({\bf q},{\bf q}^{\prime };\tau
)=-g\langle T\widetilde{\rho }_{\alpha \beta }({\bf q},\tau )\widetilde{\rho
}_{\gamma \delta }(-{\bf q}^{\prime },0)\rangle ,  \label{vdeux}
\end{equation}
where $\widetilde{\rho }_{\alpha \beta }=\rho _{\alpha \beta }-\langle \rho
_{\alpha \beta }\rangle $. By making use of the commutation relation\cite
{cote1} of the operators $\rho _{\alpha \beta }({\bf q})$ and of the HF
Hamiltonian of Eq.~(\ref{trois}), we obtain an equation of motion for the
response functions that corresponds to the HFA which we denote by $\chi ^0$.
We get (repeated spin indices are summed over)
\begin{eqnarray}
\left[ i\Omega _n+(\epsilon _\alpha -\epsilon _\beta )\right] &&\chi
_{\alpha \beta \gamma \delta }^0({\bf q},{\bf q}^{\prime };\Omega _n)=
\nonumber \\
&&\delta _{\beta ,\gamma }e^{-i\frac 12{\bf q}\times {\bf q}^{\prime }\ell
^2}\langle \rho _{\alpha \delta }({\bf q}-{\bf q}^{\prime })\rangle -\delta
_{\alpha ,\delta }e^{i\frac 12{\bf q}\times {\bf q}^{\prime }\ell ^2}\langle
\rho _{\gamma \beta }({\bf q}-{\bf q}^{\prime })\rangle  \nonumber \\
&&-\sum_{{\bf q}^{\prime \prime }}V_{\kappa \alpha }({\bf q}^{\prime \prime
}-{\bf q})e^{-i\frac 12{\bf q}\times {\bf q}^{\prime \prime }\ell ^2}\chi
_{\kappa \beta \gamma \delta }^0({\bf q}^{\prime \prime },{\bf q}^{\prime
};\Omega _n)  \nonumber \\
&&+\sum_{{\bf q}^{\prime \prime }}V_{\beta \kappa }({\bf q}^{\prime \prime }-%
{\bf q})e^{i\frac 12{\bf q}\times {\bf q}^{\prime \prime }\ell ^2}\chi
_{\alpha \kappa \gamma \delta }^0({\bf q}^{\prime \prime },{\bf q}^{\prime
},\Omega _n),  \label{vquatre}
\end{eqnarray}
where $\Omega _n$ is a boson frequency.

To calculate the response functions in the Time-Dependent Hartree-Fock
Approximation (TDHFA), and so include the correlations that give rise to
phonons and magnons, we need to sum a set of ladder and bubble diagrams\cite
{cote1}. The final equation for $\chi $ can be expressed {\it solely} in
terms the order parameters of the crystal phase!
\begin{eqnarray}
\chi _{\alpha \beta \gamma \delta }({\bf q},{\bf q}^{\prime };\Omega _n) &=&%
\widetilde{\chi }_{\alpha \beta \gamma \delta }({\bf q},{\bf q}^{\prime
};\Omega _n)  \nonumber \\
&&+\sum_{{\bf q}^{\prime \prime }}\widetilde{\chi }_{\alpha \beta \kappa
\kappa }({\bf q},{\bf q}^{\prime \prime };\Omega _n)H({\bf q}^{\prime \prime
})\chi _{\xi \xi \gamma \delta }({\bf q}^{\prime \prime },{\bf q}^{\prime
};\Omega _n),  \label{vcinq}
\end{eqnarray}
where the irreducible response function is given by
\begin{eqnarray}
\widetilde{\chi }_{\alpha \beta \gamma \delta }({\bf q},{\bf q}^{\prime
};\Omega _n) &=&\chi _{\alpha \beta \gamma \delta }^0({\bf q},{\bf q}%
^{\prime };\Omega _n)  \label{vsix} \\
&&-\sum_{{\bf q}^{\prime \prime }}\chi _{\alpha \beta \kappa \xi }^0({\bf q},%
{\bf q}^{\prime \prime };\Omega _n)X({\bf q}^{\prime \prime })\widetilde{%
\chi }_{\xi \kappa \gamma \delta }({\bf q}^{\prime \prime },{\bf q}^{\prime
};\Omega _n).  \nonumber
\end{eqnarray}
The spin and density response functions are obtained, as usual, from the
analytic continuation $i\Omega _n\rightarrow \omega +i\delta $. The
dispersion relation of the collective modes are then found by tracking the
poles of the response functions at different values of the wavevector ${\bf q%
}$ in the Brillouin zone.

In the case of a fully-polarized state, the only non-zero response functions
are $\chi _{{+--+}},\chi _{-++-}$ and $\chi _{++++}$ and so the usual spin
flip and density-density response functions are given by

\begin{equation}
\chi ^{+-}({\bf q},{\bf q}^{\prime };\Omega _n)=ge^{-q^2\ell
^2/4}e^{-q^{\prime 2}\ell ^2/4}\chi _{{+--+}}({\bf q},{\bf q}^{\prime
};\Omega _n),  \label{trente}
\end{equation}
and
\begin{eqnarray}
\chi ^{zz}({\bf q},{\bf q}^{\prime };\Omega _n) &=&\frac 14\chi ^{nn}({\bf q}%
,{\bf q}^{\prime };\Omega _n)  \label{tun} \\
&=&\frac g4e^{-q^2\ell ^2/4}e^{-q^{\prime 2}\ell ^2/4}\chi _{++++}({\bf q},%
{\bf q}^{\prime };\Omega _n).  \nonumber
\end{eqnarray}
They obey the TDHFA equations of motion

\begin{equation}
\sum_{{\bf q}^{\prime \prime }}\left[ i\Omega _n\delta _{{\bf q},{\bf q}%
^{\prime \prime }}-C_A\left( {\bf q},{\bf q}^{\prime \prime }\right)
-D_A\left( {\bf q},{\bf q}^{\prime \prime }\right) \left[ H({\bf q}^{\prime
\prime })-X({\bf q}^{\prime \prime })\right] \right] \chi _{++++}({\bf q}%
^{\prime \prime },{\bf q}^{\prime };\Omega _n)=D_A({\bf q},{\bf q}^{\prime
}),  \label{tdeux}
\end{equation}
and
\begin{equation}
\sum_{{\bf q}^{\prime \prime }}\left[ (i\Omega _n-2\Lambda )\delta _{{\bf q},%
{\bf q}^{\prime \prime }}-C_B\left( {\bf q},{\bf q}^{\prime \prime }\right)
+D_B\left( {\bf q},{\bf q}^{\prime \prime }\right) X({\bf q}^{\prime \prime
})\right] \chi _{+--+}({\bf q}^{\prime \prime },{\bf q}^{\prime };\Omega
_n)=D_B({\bf q},{\bf q}^{\prime }),  \label{ttrois}
\end{equation}
where we have defined
\begin{eqnarray}
D_A({\bf q},{\bf q}^{\prime }) &=&-2i\sin \left[ ({\bf q}\times {\bf q}%
^{\prime })\ell ^2/2\right] ,  \label{tquatre} \\
D_B({\bf q},{\bf q}^{\prime }) &=&\langle \rho _{{++}}({\bf q}-{\bf q}%
^{\prime })\rangle e^{-i({\bf q}\times {\bf q}^{\prime })\ell ^2/2},
\label{tcinq}
\end{eqnarray}
\begin{equation}
C_A({\bf q},{\bf q}^{\prime })=2i\left\langle \rho _{{++}}({\bf q}-{\bf q}%
^{\prime })\right\rangle \left[ H({\bf q}-{\bf q}^{\prime })-X({\bf q}-{\bf q%
}^{\prime })\right] \sin \left[ ({\bf q}\times {\bf q}^{\prime })\ell
^2/2\right] ,  \label{tsix}
\end{equation}
\begin{eqnarray}
C_B({\bf q},{\bf q}^{\prime }) &=&\left\langle \rho _{{++}}({\bf q}-{\bf q}%
^{\prime })\right\rangle X({\bf q}-{\bf q}^{\prime })\cos \left[ ({\bf q}%
\times {\bf q}^{\prime })\ell ^2/2\right]  \nonumber \\
&&+\left\langle \rho _{{++}}({\bf q}-{\bf q}^{\prime })\right\rangle \left[
2iH({\bf q}-{\bf q}^{\prime })-iX({\bf q}-{\bf q}^{\prime })\right] \sin
\left[ ({\bf q}\times {\bf q}^{\prime })\ell ^2/2\right] .  \label{tsept}
\end{eqnarray}
(For a Wigner crystal, ${\bf q}\rightarrow {\bf k}+{\bf G},$ ${\bf q}%
^{\prime }\rightarrow {\bf k}+{\bf G}^{\prime }$ etc. where ${\bf k}$ is a
vector restricted to the first Brillouin zone of the crystal.) The problem
of calculating the spin-flip and density-density response functions is then
reduced to a matrix-diagonalization problem. The two response functions
decouple. The matrix eigenvalues are the collective excitations associated
with the two response functions, phonons in the case of $\chi ^{nn}$ and
magnons in the case of $\chi ^{+-}$.

\section{Hartree Fock Approximation for the Ground State}

We first apply the above formalism to examine the nature of the magnetic
order in the Wigner crystal ground state. In the Hartree-Fock approximation
the ground state at strong magnetic fields always has broken translational
symmetry.\cite{earlyhf} This result of the Hartree-Fock approximation is an
artifact. As we mentioned in the introduction, the true ground state has
broken translational symmetry only\cite{skyrm} for $\nu <0.23$.
Nevertheless, as we discuss further below the Hartree-Fock approximation
does describes {\it the ground state}, reasonably accurately when the ground
state {\it is} a Wigner crystal. Of course the Hartree-Fock approximation
completely misrepresents the excitation spectrum of the Wigner crystal,
since it misses the phonon and magnon collective modes captured by the
time-dependent Hartree-Fock approximation.

In two-dimensions, Coulomb interactions favor\cite{mara} a triangular
lattice for the Wigner crystal. We will find that the energy scale
associated with magnetic order is much smaller than the Coulomb energy
scale. We therefore expect the structure of the Wigner crystal to be the
triangular lattice structure dictated by Coulomb interactions. We also
expect that the interactions between the spins on the triangular lattice
sites will be predominantly nearest neighbor since the overlap between wave
functions on different sites is quite small in a strong magnetic field. We
check this approximation below by comparing the dispersion relation of the
spin waves in the TDHFA with that given by a Heisenberg model with only
nearest-neighbour exchange coupling. The ground state for two-dimensional
spin $1/2$ particles with nearest-neighbour interactions on a triangular
lattice is expected to have long range order for both ferromagnetic and
antiferromagnetic interactions.\cite{tslaf} However, because of frustration,
the order is rather subtle for the antiferromagnetic case. (The triangular
lattice is {\bf not} a bipartite lattice.) Our primary objective in this
subsection is to determine whether the interactions is ferromagnetic or
antiferromagnetic by comparing the energy of these two states. For that
purpose it is more useful to consider the case of two-dimensional electrons
on a square lattice since it is bipartite and both antiferromagnetic and
ferromagnetic states have a simple structure. We do so even though the
ground state of the two-dimensional electron solid does not occur in this
structure.

\subsection{Maki-Zotos Wavefunction}

It is instructive to begin by generalizing the wavefunction for spinless
electrons employed by Maki and Zotos\cite{maki} in their study of the strong
field Wigner crystal. We define
\begin{equation}
\Psi =(N!)^{-1/2}\det \left| \psi _{{\bf R}_j}({\bf r}_i)\chi _{{\bf n}%
_j}^i\right| .  \label{eq:mz}
\end{equation}
Here ${\bf R}_j$ is the $j$-th lattice vector,
\begin{equation}
\psi _{{\bf R}}({\bf r})=\frac 1{\sqrt{2\pi \ell ^2}}\exp \left( \frac{-|%
{\bf r}-{\bf R}|^2-2i(xR_y-yR_x)}{4\ell ^2}\right) ,  \label{eq:spwf}
\end{equation}
is the lowest Landau level wavefunction\cite{gauge} for an electron whose
quantized cyclotron orbit is centered on ${\bf R}$, and $\chi _{{\bf n}%
}=(\cos (\theta /2),\sin (\theta /2)\exp (i\phi ))$ is a spinor oriented in
the ${\bf n}=(\sin (\theta )\cos (\phi ),\sin (\theta )\sin (\phi ),\cos
(\theta ))$ direction. In this wavefunction the cyclotron orbits of
electrons near different lattice sites are uncorrelated and the electron
spin orientation at a given lattice site is arbitrary. In the range of $\nu $
where the ground state is a Wigner crystal it is an excellent approximation%
\cite{maki,overlpcaveat} to ignore the lack of orthogonality between
cyclotron orbits centered at different lattice sites. Making this
approximation, it is easy to derive an expression for
\begin{equation}
E\equiv \frac{\langle \Psi |\sum_{i<j}e^2\left| {\bf r}_i-{\bf r}_j\right|
^{-1}|\Psi \rangle }{\langle \Psi |\Psi \rangle }.  \label{eq:vareng}
\end{equation}
In Eq.~\ref{eq:vareng}, we have dropped the Zeeman energy which can easily
be added if the electronic $g$-factor is non-zero. The fact that the kinetic
energy, taken as the zero of energy above, is quantized is important in
determining the favored magnetic order. Following Maki and Zotos we find
that
\begin{equation}
E={\frac 12}\sum_{i\ne j}\left[ I\left( \left| {\bf R}_j-{\bf R}_i\right|
\right) -\left( \frac{1+{\bf n}_i\cdot {\bf n}_j}2\right) J_{MZ}\left(
\left| {\bf R}_j-{\bf R}_i\right| \right) \right] ,  \label{eq:mzeng}
\end{equation}
where
\begin{equation}
I(R)=\left( \frac{e^2}\ell \right) \frac{\sqrt{\pi }}2\exp \left( -R^2/8\ell
^2\right) I_0\left( R^2/8\ell ^2\right) ,  \label{eq:direct}
\end{equation}
and
\begin{equation}
J_{MZ}(R)=\exp (-R^2/4\ell ^2)I(R),  \label{eq:exchange}
\end{equation}
In this equation $I(R)$ and $J(R)$ are respectively the direct and exchange
two-body matrix elements of the Coulomb interaction for lowest-Landau-level
cyclotron orbits whose centers are separated by ${\bf R}$. The explicit
expression for the matrix element in the exchange term, which is sensitive
to the relative orientations of the spins on the two sites, is
\begin{equation}
J_{MZ}\left( \left| {\bf R}_j-{\bf R}_i\right| \right) =e^2\int d{\bf r}\int
d{\bf r}^{\prime }\frac{\psi _{{\bf R}_j}^{*}({\bf r})\psi _{{\bf R}_j}({\bf %
r}^{\prime })\psi _{{\bf R}_i}^{*}({\bf r}^{\prime })\psi _{{\bf R}_i}({\bf r%
})}{\left| {\bf r}-{\bf r}^{\prime }\right| }.  \label{def}
\end{equation}
For $R\gg \ell $,
\begin{equation}
J_{MZ}(R)\approx \frac{e^2}R\exp \left( -R^2/4\ell ^2\right) .
\label{eq:appexchange}
\end{equation}
This approximate expression for $J_{MZ}(R)$ is accurate to better than $5\%$
even for neighboring sites over the range of Landau level filling factors
where the ground state is a Wigner crystal.

It is evident from Eq.(\ref{eq:mzeng}) that if the ground state is
approximated by the Maki-Zotos wave function, a ferromagnetic state in which
all spins are parallel will be energetically favored. The energy increase
when the relative orientation of spins on two sites separated by $R$ changes
from parallel to antiparallel is $J(R)$. For similar
single-Slater-determinant variational wave functions at zero magnetic field,
the tendency would be to favor antiferromagnetic orientations on neighboring
sites\cite{thouless,ceperley,louie} except possibly when multi-site ring
exchanges become important.
(Multi-site ring exchanges are less important for the strong magnetic field
Wigner crystal because magnetic confinement results in orbitals which are
more strongly localized around lattice sites.) In the weak field case,
having opposite spins on neighboring sites reduces the kinetic energy
density required by the Pauli exclusion principle in the region between the
sites. In the strong magnetic field limit, the kinetic energy is quantized
and is independent of the spin-configuration so this mechanism favoring
antiferromagnetism is not operative. Nevertheless the Maki-Zotos
wavefunction is a single-Slater-determinant and conclusions based upon its
use should be examined critically. It is known, for example, that
correlations can result in spin-singlet fluid ground states\cite{rasolt}
whereas the Hartree-Fock approximation would always predict ferromagnetic
ground states. At zero magnetic field, the contribution from low-energy,
long-wavelength phonon modes to the zero-point motion gives rise to
long-range correlations which, for example, make the static structure factor
vanish more quickly ($\propto q^{3/2}$) than it would for a system with
short range interactions. At strong magnetic field, even stronger
correlations which make the static structure factor vanish as $q^2$ result
from the contribution to the zero-point motion of the collective cyclotron
mode of all electrons. (In a Jastrow-Slater variational wavefunction such as
that used by Zhu and Louie\cite{louie} the correlation factors would have to
have a logarithmic spatial dependence in order to capture the correct
long-distance ground-state correlations.) We cannot completely rule out on
the basis of our calculations the possibility that correlations could
invalidate our conclusion that the ground state is ferromagnetic. However we
consider this to be extremely unlikely.

\subsection{Self-Consistent Hartree-Fock Calculations}

One possible mechanism in favor of antiferromagnetism is the possibility of
spreading the charge associated with a given lattice site more widely in the
case of antiferromagnetic configurations which could reduce the
electrostatic energy. To probe the competition a little more deeply we have
performed self-consistent Hartree-Fock calculations, based on the formalism
of the previous section, comparing the energy of ferromagnetic and
antiferromagnetic states on a square lattice. We now discuss the results of
these calculations.

In the Hartree-Fock approximation, the spin-order is unidirectional on a
square lattice for both ferromagnetic and antiferromagnetic interactions. We
choose a spin-quantization axis which is along the direction of the Zeeman
coupling if one is present and is otherwise arbitrary. This allows us to set
the order parameters which are off-diagonal in the spin indices to zero and
simplify our calculation. Let $a_0$ be the lattice constant of the
ferromagnetic square lattice with density $n=1/a_0^2$ such that $2\pi n\ell
^2=\nu $. In the antiferromagnetic case we assume that the spin density is
oppositely directed on the two sublattices (which have lattice constant $%
\sqrt{2}a_0,$ and have a relative shift of ${\bf a=}\sqrt{2}\left( \frac 12,%
\frac 12\right) a_0$) so that
\begin{equation}
\left\langle \rho _{{--}}({\bf G})\right\rangle =e^{-i{\bf G}\cdot {\bf a}%
}\left\langle \rho _{{++}}({\bf G})\right\rangle  \label{quarante}
\end{equation}
where ${\bf G}$ is a {\em sublattice} reciprocal lattice vector (with
modulus $\left| {\bf G}\right| =2\pi /\sqrt{2}a_0$). (We choose our
coordinate system so that the primitive lattice vectors of the sublattice
are along the Cartesian axes.) Eq. (\ref{dixp}) can again be simplified to a
single equation
\begin{equation}
\left( i\omega _n+\mu \right) G_{{++}}({\bf G},\omega _n)-\sum_{{\bf G}%
^{\prime }}\widetilde{F}({\bf G},{\bf G}^{\prime })G_{{++}}({\bf G}^{\prime
},\omega _n)=\delta _{{\bf G},0},  \label{qq}
\end{equation}
where
\begin{equation}
\widetilde{F}({\bf G},{\bf G}^{\prime })=\exp \left[ \frac 12i{\bf G}\times
{\bf G}^{\prime }\ell ^2\right] \left[ \left( 1+e^{-i{\bf G}\cdot {\bf a}%
}\right) H({\bf G})-X({\bf G})\right] \left\langle \rho _{{++}}(-{\bf G}%
)\right\rangle .  \label{qqq}
\end{equation}
The ground-state energy per particle becomes
\begin{equation}
E_{{+-}}=\frac 1\nu \sum_{{\bf G}}\left[ H({\bf G})\left( 1+\cos ({\bf G}%
\cdot {\bf a})\right) -X({\bf G})\right] \left| \left\langle \rho _{{++}}(%
{\bf G})\right\rangle \right| ^2.  \label{qdeux}
\end{equation}

We have solved these equations self-consistently. Because of the variational
nature of the Hartree-Fock approximation, these solutions provide us with
the lowest energy single-Slater determinant consistent with the assummed
magnetic and translational broken symmetry. In particular, the solutions to
these equations will always give a lower energy than the energy for the
corresponding Maki-Zotos wavefunction. The optimization process implicit in
obtaining a self-consistent solution of the Hartree-Fock equations results
in cyclotron orbits on each lattice site which are distorted by their
average environments, including their magnetic environments, in a way which
minimizes the total interaction energy. It is still true, however, that the
cyclotron orbits on different sites are not correlated with each other. The
error introduced as a consequence can be estimated by using a harmonic
approximation for the strong field Wigner crystal, which is reasonably
accurate from an energetic point of view throughout the regime where the
ground state is an electron crystal. In the harmonic approximation the
many-body Schrodinger equation can be solved exactly and the ground-state
energy is the sum of the classical Madelung energy and the quantum
zero-point energy i.e.
\begin{equation}
E_{harmonic}=-0.78213\nu ^{1/2}+0.24101\nu ^{3/2},  \label{eharmo}
\end{equation}
for the hexagonal lattice. The Hartree-Fock approximation describes the
Madelung term exactly (in the limit $\nu \rightarrow 0,$ the HFA energy
coincides with the classical energy of a point lattice) and overestimates%
\cite{ahmaustria} the zero-point energy by approximately $25\%$ (at $\nu
=0.2 $ ).

The results of our calculations are summarized in Tables 1and 2. In Table 1,
we list the ground-state energy per electron in the HFA for the square
lattice antiferromagnetic (SLA) and ferromagnetic (SLF) states as well as
for the triangular lattice ferromagnetic state (TLF). Table 2 contains a
similar calculation{\bf \ }using a simplified form of the Maki and Zotos
wave function where we have neglected the overlapping between two wave
functions centered on different sites so that the single-electron density
can be approximated by
\begin{equation}
\left\langle n({\bf r})\right\rangle =\sum_i\ \left| \psi _{{\bf R}_i}({\bf r%
})\right| ^2=\frac 1{2\pi \ell ^2}\sum_ie^{-({\bf r}-{\bf R}_i)^2/2\ell ^2},
\label{ajout1}
\end{equation}
or, equivalently, in the ferromagnetic case
\begin{equation}
\left\langle \rho ({\bf G})\right\rangle _{MZ}=\nu \ e^{-G^2\ell ^2/4}.
\label{ajout2}
\end{equation}
(For the SLA case, $\nu \rightarrow \nu /2$ and the ${\bf G}^{\prime }s$ are
replaced by the sublattices reciprocal lattice vectors.) We use the order
parameters defined by Eq.(\ref{ajout2}) in Eq.(\ref{vneuf}) to compute the
Maki-Zotos ground-state energies tabulated in Table II. We remark that this
procedure is exactly equivalent to computing Eq.(\ref{eq:mzeng}) (when the
interaction with a positive homogeneous background of charges is added to
this last equation). Note also that these results include only the Coulomb
energy.
These Tables report also the results of calculations performed for filling
factors where the ground state is {\it not} believed to be a Wigner crystal.
These large $\nu $ results are intended to illustrate trends in the
self-consistent Hartree-Fock approximation solutions and not to be
physically realistic.\cite{note1}

We see that for the larger filling factors, the difference in energy between
Hartree-Fock square lattice ferromagnetic and antiferromagnetic states, $%
\Delta E_{spin}=(E_{SLF}-E_{SLA})/E_{SLF}$ agrees quite closely with what
would be predicted by the Maki-Zotos wavefunction (Eq. (\ref{ajout2})). At
smaller filling factors, however, the Hartree-Fock energy difference between
these two spin states is much bigger than what would be predicted by the
Maki-Zotos wavefunction. The energy reduction due to the added variational
freedom compared to the Maki-Zotos wavefunction is larger for the
ferromagnetic state than for the antiferromagnetic state and this leads to
an increased energy difference between the two states. (See also Table 3
where, as discussed below, $J_{MZ}$ is proportional to $(E_{SLF}-E_{SLA})$
evaluated with the Maki-Zotos wavefunction and $J_{HFA}$ is proportional to $%
(E_{SLF}-E_{SLA})$ evaluated in the HFA.)\cite{note2} In both the
Makis-Zotos and HF approximations, the ferromagnetic state has the lowest
energy. In Fig. 1, we plot the {\it difference} in density: $\left\langle n(%
{\bf r})\right\rangle _{HFA}-\left\langle n({\bf r})\right\rangle _{MZ}$ for
the SLF and SLA at filling factor $\nu =1/8.$ It is clear from this figure
that the HFA minimizes the Coulomb energy in both the SLF and SLA cases by
removing charges along the direction of the nearest-neighbor sites and
putting them along the direction of the next-nearest-neighbor sites. This is
just what we expect at such a small filling factor where overlap between
wave functions on different sites is very small and the ground-state energy
is dominated by the (direct) Coulomb interaction. According to our
calculations more charge redistribution occurs in the ferromagnetic case.
For the triangular lattice, a similar calculation gives a much smaller
difference in densities reflecting a loss in the variational freedom due to
the higher coordination number of the triangular lattice.

We remark that similar self-consistent calculations for a single-band
Hubbard model at half-filling would find the antiferromagnetic state to be
lower in energy, correctly reflecting the superexchange coupling in that
system.\cite{vignale} We also see that, as anticipated above, the difference
between the square lattice ferromagnetic state and the triangular lattice
ferromagnetic state energy $\Delta E_{Coulomb}=(E_{TLF}-E_{SLF})/E_{TLF}$ is
much larger than the difference between ferromagnetic and antiferromagnetic
states on the same lattice. This energy difference is almost constant over
the range of filling factors considered here, decreasing slowly as $\nu $
decreases. For $\nu \to 0,\ \Delta E_{Coulomb}$ approaches its Madelung
energy value,\cite{mara} $0.53\%$.

\section{Collective Mode Calculations}

As we described above, the ground-state order parameters can be used to
calculate the spin-wave collective modes. In the ferromagnetic ground state,
we showed that the density response function, $\chi ^{nn},$ or equivalently
the longitudinal spin response function, $\chi ^{zz}=\chi ^{nn}/4,$ are
uncoupled from the transverse spin response function $\chi ^{+-}$ . The
poles of the longitudinal spin response function are nothing but the phonons
of the Wigner crystal for which we have already computed the dispersion
relation in Ref.~\cite{cote1}. The transverse spin excitations are the
magnon collective modes of the ferromagnetic Wigner crystal. Figs.~1 and 2
show the TDHFA\ dispersion relations for the SF and TF lattices,
respectively, at different values of the filling factor. (When the
electronic g-factor is non-zero all magnon collective modes energies are
increased by the Zeeman gap $g^{*}\mu _BB$.
In the simple colinear states (SLF, SLA and TLF) that we consider here, a
small Zeeman term has no effect on the calculation of the order parameters.)
We remark that, because of the numerical approach used in this work, it is
not possible to obtain the dispersion relations at very small filling
factors without having to consider a prohibitively large number of
reciprocal lattice vectors. For the square lattice, we were not able to
obtain accurate results for $\nu <1/5$ while for the triangular lattice $\nu
=1/7$ was the lower limit.

It is interesting to compare these magnon dispersion relations with the
dispersion relation of the spin waves of the Heisenberg model where the
spins are localized on the lattice sites and the Hamiltonian is given by
\begin{equation}
H=-J\sum_{i{\bf \delta }}{\bf S}_i\cdot {\bf S}_{i+{\bf \delta }}.
\label{cinquante}
\end{equation}
(The summation is over the lattices sites $i$ and the $\nu _o$ nearest
neighbours of the lattice. Note that this convention for the exchange
constant results in double-counting each neighbour pair.) In that case,
\begin{equation}
\omega ({\bf k})=2J\nu _os(1-\gamma _{{\bf k}}),  \label{cun}
\end{equation}
with
\begin{equation}
\gamma _{{\bf k}}=\frac 1{\nu _o}\sum_{{\bf \delta }}e^{i{\bf k}\cdot {\bf R}%
_\delta },  \label{cdeux}
\end{equation}
and $s=1/2$.

The solid lines in Figs.~1 and 2, show the dispersion relation obtained from
a nearest-neighbor interaction Heisenberg model with the interaction
strength chosen to reproduce the TDHFA numerical results. The fit is quite
good and becomes almost perfect at smaller filling factors. (The discrepancy
at $\nu =1/3$ can be improved by fitting with non-zero
next-nearest-neighbour coupling.) The exchange integral $J_{TDHFA}$
obtained, in this way, from the TDHFA dispersion relation, for different
filling factors, is listed in Table~3 (square lattice) and Table 4
(triangular lattice) . These tables also show values of the exchange
integral calculated in two other ways. From Table~1, we see that the
ground-state energy is minimal for the polarized square lattice. We can
estimate the strength of the exchange coupling from this energy difference
as follows. We assume that, as far as the magnetic degrees of freedom are
concerned, the Hartree-Fock solution yields an Ising approximation to the
antiferromagnetic ground state, {\it i.e.} it does note capture the quantum
fluctuations which would be present in a true antiferromagnetic ground
state. It is then easy to see\cite{madelung} that, for the square lattice
\begin{equation}
J_{HFA}=\frac{E_{SLA}-E_{SLF}}2.  \label{ctrois}
\end{equation}
This expression assumes that non-nearest neighbour exchange coupling is
negligible. (Note that a similar calculation is not possible for the
triangular lattice because of frustration.) We can also compute the exchange
integral directly from the Maki-Zotos wavefunction expression, Eq.(\ref{def}%
). This gives Eq.(\ref{eq:direct}) or, using $2\pi n\ell ^2=\nu $ with $%
n=1/\alpha a_0^2$ ($\alpha =1$ (SF) or $\alpha =\sqrt{3}/2$ (TF))
\begin{equation}
J_{MZ}=\left( \frac{e^2}\ell \right) \left( \frac \pi 4\right) ^{1/2}e^{-%
\frac{3\pi }{4\alpha \nu }}I_0\left( \frac \pi {4\alpha \nu }\right) .
\label{j3}
\end{equation}
Note that if the Maki-Zotos energies are used in Eq. (\ref{ctrois}) instead
of the HFA energies, $J_{MZ}$ is recovered exactly as long as
non-nearest-neighbour interactions are negligible.

For the TLF, the exchange integral obtained from the TDHFA is only slightly
smaller than $J_{MZ}$. For the square lattice, the exchange integral
obtained from the TDHFA is smaller than both $J_{MZ}$ and the HFA value over
the range of $\nu$ where we are able to complete calculations.
At smaller filling factors, since the difference between the two spin states
decreases faster with the Maki-Zotos wavefunction than with the HFA, the HF
value of $J$ is much larger than the Maki-Zotos value. We expect the TDHFA
result to remain close to the HFA result in this regime. Unfortunately, we
cannot check this assumption numerically since we cannot compute the TDHFA
value of $J$ at smaller filling factor (the matrix size becomes
prohibitively large). However, since the TDHFA is obtained from a functional
differentiation of the HFA, our assumption seems reasonable. In any case,
the present result shows clearly that, in the Wigner crystal, the
interaction between spins, at small filling factor, is mainly from
nearest-neighbors.

In the small wave vector limit, the Heisenberg dispersion relation on the
triangular lattice is given by $\omega (k)=\frac 32J\left( ka_0\right)
^2\equiv D(k\ell )^2$ where $D=2\pi \sqrt{3}J/\nu $ in units of $e^2/\ell .$
With $J$ given by $J_{TDHFA}$ as calculated above, we find that $D=0.024$ at
$\nu =1/3$ and $D=9.3$ $\times $ $10^{-3}$ at $\nu =1/5.$ These values
should be compared with those of the liquid state where $D=4\pi \ell ^2\rho
_s/\nu $ where $\rho _s$ is the spin-stiffness. For the liquid state it is
possible to express the spin-stiffness in terms of the pair correlation
function\cite{rasolt,moon} and this has been evaluated using an
hypernetted-chain-approximation for the liquid state pair-correlation
function in Ref. \cite{moon}. For the liquid state we find that, $%
D=0.035(e^2/\ell )$ at $\nu =1/3$ and $D=0.015(e^2/\ell )$ at $\nu =1/5.$ We
see that the spin-stiffness is larger for the liquid state and increasingly
so as the filling factor decreases. This result is consistent with the view
of the strongly correlated electron states as quantum melted crystals of
electrons whose size is smeared on a magnetic length scale by rapid
cyclotron motion. When long-range order is lost, the cyclotron orbits will
overlap more strongly on average and the relative spin-orientation of nearby
electrons will assume a larger importance.

\section{SUMMARY}

In the strong-magnetic-field limit, we have argued that the lowest-energy
spin state of the Wigner crystal is the ferromagnetic state. Our conclusion
is based in part on a comparison of ferromagnetic and antiferromagnetic
state energies of square-lattice Wigner crystal states calculated in the
Hartree-Fock approximation and in part on the observation that the
superexchange mechanism, which tends to favor antiferromagnetism, is absent
in the strong magnetic field limit. The spin-wave dispersion relations,
which we compute in the TDHFA, show that the ferromagnetic lattice is stable
at filling factors where crystallization occurs. In this limit, the
interactions between spins on the lattice are dominated by nearest-neighbor
exchange coupling. Our results appear to show that small distortions of the
wavefunctions for electrons on one lattice site, due to their interactions
with electrons on nearby lattice sites, are responsible for a large relative
increase in the small exchange couplings at small filling factors. A
comparison with the liquid state at filling factors $\nu =1/3$ and $\nu =1/5$
shows that the spin-stiffness of the ferromagnetic liquid states which occur
at these filling is substantially larger than that of corresponding crystal
states. In closing we remark that the recent sucessful application\cite
{barrett} of nuclear-magnetic-resonance methods to two-dimensional electron
system, suggests that the magnetic properties of two-dimensional electron
systems in the regime where the Wigner crystal state occurs will soon be
open to experimental investigation so that our conclusions can be tested.
These experiments should open up a host of interesting new questions,
related to disorder and thermal fluctuations.

\section{Acknowledgments}

This research was supported in part by NSF grant DMR-9416906, and by the
Natural Sciences and Engineering Research Council of Canada (NSERC) and the
Fonds pour la formation de chercheurs et l'aide \`a la recherche (FCAR) from
the Government of Qu\'ebec.

\newpage\

\begin{center}
TABLE CAPTIONS
\end{center}

\bigskip\

\begin{description}
\item[Table 1]  Energy of the square and triangular ferromagnetic states
(SLF,TLF) and of the antiferromagnetic state on the square lattice (SLA) in
the Hartree-Fock approximation in units of $e^2/\ell .$ The relative energy
difference between the ferromagnetic and antiferromagnetic states on the
square lattice is given by $\Delta E_{spin}$ while the relative difference
in energy between the ferromagnetic state on the triangular and square
lattice is given by $\Delta E_{Coulomb}.$

\item[Table 2]  Energy of the square and triangular ferromagnetic states
(SLF,TLF) and of the antiferromagnetic state on the square lattice (SLA) in
the Maki-Zotos approximation in units of $e^2/\ell .$ The relative energy
difference between the ferromagnetic and antiferromagnetic states on the
square lattice is given by $\Delta E_{spin}$ while the relative difference
in energy between the ferromagnetic state on the triangular and square
lattice is given by $\Delta E_{Coulomb}.$

\item[Table 3]  Value of the exchange integral, in units of $e^2/\ell ,$ and
on the square lattice obtained from various approximations:\ $J_{MZ}$ is
from the definition of the exchange integral given in Eq. (\ref{j3}), $%
J_{HFA}$ is obtained from the energy difference between the ferromagnetic
and antiferromagnetic states (Eq. (\ref{ctrois})), and $J_{TDHFA}$ is
obtained by fitting the TDHFA spin-wave dispersion relation with the
spin-wave dispersion relation of the Heisenberg model (see text).

\item[Table 4]  Value of the exchange integral , in units of $e^2/\ell ,$
and on the triangular lattice obtained from the various approximations
described in Table 3.
\end{description}

\newpage\

\begin{center}
FIGURE CAPTIONS
\end{center}

\bigskip\

\begin{description}
\item[Fig. 1]  Difference in densities between the Hartree-Fock and
Maki-Zotos approximations at filling factor $\nu =1/8$ and in units of $%
1/a_0^2$ for (a) the ferromagnetic square lattice and (b) antiferromagnetic
square lattice.
For the ferromagnetic state, the lattice sites are indicated by gray
circles. For the antiferromagnetic state, sites on one sublattice are
indicated by black circles and sites on the other sublattice are indicated
by empty squares. Note that the orientation of the lattice differs in (a)
and (b).
\end{description}

\begin{itemize}
\item[Fig. 2]  Dispersion relation of the spin waves of the ferromagnetic
square lattice obtained from the TDHFA and from the Heisenberg model at
different filling factors. The dispersion relation is plotted along the
edges of the irreducible Brillouin zone of the square lattice with lattice
spacing $a.$ In units of $2\pi /a,$ $\Gamma =(0,0),$ $J=(1/2,1/2),$ $%
X=(1/2,0).$
\end{itemize}

\begin{description}
\item[Fig. 3]  Dispersion relation of the spin waves of the ferromagnetic
triangular lattice obtained from the TDHFA and from the Heisenberg model at
different filling factors. The dispersion relation is plotted along the
edges of the irreducible Brillouin zone of the triangular lattice with
lattice spacing $a.$ In units of $2\pi /a,$ $\Gamma =(0,0),$ $J=(1/3,1/\sqrt{%
3}),$ $X=(1/\sqrt{3},0).$
\end{description}

\newpage\

\bigskip\ \bigskip\

\bigskip\ \bigskip\

\begin{center}
\begin{tabular}{|c|c|c|c|c|c|}
\hline
$\nu $ & $E_{SLF}$ & $E_{SLA}$ & $E_{TLF}$ & $\Delta E_{spin}$ (\%) & $%
\Delta E_{Coulomb}$ (\%) \\ \hline
$1/3$ & -.3857672 & -.3823134 & -.3884928 & .90 & .70 \\ \hline
$1/4$ & -.3484399 & -.3478963 & -.3511452 & .16 & .77 \\ \hline
$1/5$ & -.3196321 & -.3195124 & -.3219969 & .037 & .73 \\ \hline
$1/6$ & -.2964916 & -.2964471 & -.2985717 & .015 & .70 \\ \hline
$1/7$ & -.2775104 & -.2774874 & -.2793787 & .0083 & .67 \\ \hline
$1/8$ & -.2616473 & -.2616346 & -.2633555 & .0049 & .65 \\ \hline
$1/10$ & -.2365415 & -.2365379 & -.2380218 & .0015 & .62 \\ \hline
\end{tabular}
\end{center}

\bigskip\
\bigskip\
\bigskip\

R. C\^{o}t\'{e} and A.H. MacDonald, Table 1

\newpage\

\bigskip\ \bigskip\

\bigskip\ \bigskip\

\begin{center}
\begin{tabular}{|c|c|c|c|c|c|}
\hline
$\nu $ & $E_{SLF}$ & $E_{SLA}$ & $E_{TLF}$ & $\Delta E_{spin}$ (\%) & $%
\Delta E_{Coulomb}$ (\%) \\ \hline
$1/3$ & -.3858614 & -.3814086 & -.3885208 & 1.15 & .68 \\ \hline
$1/4$ & -.3482640 & -.3474804 & -.3511413 & .23 & .82 \\ \hline
$1/5$ & -.3194893 & -.3193454 & -.3219900 & .045 & .78 \\ \hline
$1/6$ & -.2964083 & -.2963812 & -.2985680 & .0091 & .72 \\ \hline
$1/7$ & -.2774646 & -.2774594 & -.2793770 & .0019 & .68 \\ \hline
$1/8$ & -.2616223 & -.2616213 & -.2633548 & .00038 & .66 \\ \hline
$1/10$ & -.2365341 & -.2365340 & -.2380216 & .000016 & .63 \\ \hline
\end{tabular}
\end{center}

\bigskip\

\bigskip\

\bigskip\

\bigskip\

R. C\^{o}t\'{e} and A.H. MacDonald, Table 2

\newpage\

\bigskip\ \bigskip\

\bigskip\ \bigskip\

\begin{center}
\begin{tabular}{|c|c|c|c|}
\hline
$\nu $ & $J_{MZ}$ & $J_{HFA}$ & $J_{TDHFA}$ \\ \hline
$1/3$ & .22 $\times $ 10$^{-2}$ & .17 $\times $ 10$^{-2}$ & .13 $\times $ 10$%
^{-2}$ \\ \hline
$1/4$ & .39 $\times $ 10$^{-3}$ & .27 $\times $ 10$^{-3}$ & .24 $\times $ 10$%
^{-3}$ \\ \hline
$1/5$ & .72 $\times $ 10$^{-4}$ & .60 $\times $ 10$^{-4}$ & .43 $\times $ 10$%
^{-4}$ \\ \hline
$1/6$ & .14 $\times $ 10$^{-4}$ & .22 $\times $ 10$^{-4}$ & -- \\ \hline
$1/7$ & .26 $\times $ 10$^{-5}$ & .11 $\times $ 10$^{-4}$ & -- \\ \hline
$1/8$ & .50 $\times $ 10$^{-6}$ & .63 $\times $ 10$^{-5}$ & -- \\ \hline
$1/10$ & .19 $\times $ 10$^{-7}$ & .18 $\times $ 10$^{-5}$ & -- \\ \hline
\end{tabular}
\end{center}

\bigskip\

\bigskip\

\bigskip\

\bigskip\

R. C\^{o}t\'{e} and A.H. MacDonald, Table 3

\newpage\

\bigskip\ \bigskip\

\bigskip\ \bigskip\

\begin{center}
\begin{tabular}{|c|c|c|}
\hline
$\nu $ & $J_{MZ}$ & $J_{TDHFA}$ \\ \hline
$1/3$ & .99 $\times $ 10$^{-3}$ & .74 $\times $ 10$^{-3}$ \\ \hline
$1/4$ & .14 $\times $ 10$^{-3}$ & .12 $\times $ 10$^{-3}$ \\ \hline
$1/5$ & .20 $\times $ 10$^{-4}$ & .17 $\times $ 10$^{-4}$ \\ \hline
$1/6$ & .29 $\times $ 10$^{-5}$ & .25 $\times $ 10$^{-5}$ \\ \hline
$1/7$ & .44 $\times $ 10$^{-6}$ & .40 $\times $ 10$^{-6}$ \\ \hline
$1/8$ & .67 $\times $ 10$^{-7}$ & -- \\ \hline
$1/10$ & .16 $\times $ 10$^{-8}$ & -- \\ \hline
\end{tabular}
\end{center}

\bigskip\

\bigskip\

\bigskip\

\bigskip\

R. C\^{o}t\'{e} and A.H. MacDonald, Table 4


\begin{references}
\bibitem{fieldcaveat}  We will imagine that a magnetic field, if present,
couples only to the orbital electronic degrees of freedom. This situation is
frequently close to being realized for electrons in semiconductors where the
$g$-factor is often small.
\bibitem{gauge}  We choose the symmetric gauge here to facilitate
comparisons with the work of Maki and Zotos.
\bibitem{overlpcaveat}  The non-orthogonality correction to the exchange
energy for electrons centered by R is $\propto \exp (-R^2/2\ell ^2)$ while
the exchange energy is $\propto \exp (-R^2/4\ell ^2)$.
\bibitem{note2}  We note that, in computing $\Delta E_{spin}$ for $\nu =1/10,
$ we have kept more than the 7 digits indicated in Tables 1 and 2.
\end{references}
\end{document}